\documentclass[runningheads]{llncs}
\usepackage[T1]{fontenc}
\usepackage{graphicx}

\usepackage{listings}
\usepackage{xcolor}

\usepackage{amsmath}

\usepackage{subcaption}
\usepackage{hyperref}

\usepackage{color}

\usepackage{bbding}

\begin{document}
\title{ColBERT-serve: Efficient Multi-Stage Memory-Mapped Scoring}

\authorrunning{K. Huang et al.}

\author{Kaili Huang\Envelope\inst{1}\textsuperscript{\dag}\and
Thejas Venkatesh\inst{2}\textsuperscript{\dag}\and
Uma Dingankar\inst{3}\textsuperscript{\dag}\textsuperscript{\ddag}\and
Antonio Mallia\inst{4} \and
Daniel Campos\inst{5} \and
Jian Jiao\inst{1} \and
Christopher Potts\inst{6} \and
Matei Zaharia\inst{7} \and
Kwabena Boahen\inst{6} \and
Omar Khattab\inst{6} \and
Saarthak Sarup\inst{6} \and
Keshav Santhanam\inst{6}}

\institute{
Microsoft, Redmond, WA, USA\\
\email{kaili.khuang@gmail.com, jian.jiao@microsoft.com} \and
Samaya AI, Mountain View, CA, USA\\
\email{thejas@stanford.edu} \and
Foundry, Palo Alto, CA, USA\\
\email{uma@mlfoundry.com} \and
Pinecone, New York, NY, USA\\
\email{me@antoniomallia.it} \and
Snowflake, New York, NY, USA\\
\email{daniel.campos@snowflake.com} \and
Stanford University, Stanford, CA, USA\\
\email{\{cgpotts, boahen, okhattab, ssarup, keshav2\}@stanford.edu} \and
UC Berkeley, Berkeley, CA, USA\\
\email{matei@berkeley.edu}}
\maketitle 

\renewcommand{\thefootnote}{} % Remove footnote numbering
\footnotetext{\textsuperscript{\dag}K. Huang, T. Venkatesh, and U. Dingankar contributed equally to this work.}
\footnotetext{\textsuperscript{\ddag}Work by U. Dingankar was done while at Stanford.}
\renewcommand{\thefootnote}{\arabic{footnote}} % Re-enable numbering
\setcounter{footnote}{0} % Reset counter 

\begin{abstract}
We study serving retrieval models, specifically late interaction models like ColBERT, to many concurrent users at once and under a small budget, in which the index may not fit in memory. We present ColBERT-serve, a novel serving system that applies a memory-mapping strategy to the ColBERT index, reducing RAM usage by 90\% and permitting its deployment on cheap servers, and incorporates a multi-stage architecture with hybrid scoring, reducing ColBERT's query latency and supporting many concurrent queries in parallel.
\end{abstract}
\section{Introduction}
Multi-vector late-interaction retrievers like ColBERT~\cite{khattab2020colbert} and ColPali~\cite{faysse2024colpaliefficientdocumentretrieval} have demonstrated state-of-the-art quality and superior generalization \cite{thakur2021beir} while maintaining low latency, but despite major progress in compressing their embeddings \cite{santhanam2022colbertv2,hofstatter2022colberter}, hosting a ColBERT index of Wikipedia (20M passages) via PLAID \cite{santhanam2022plaid} demands nearly 100GB of RAM. This poses a challenge for serving such models on cheap servers with little RAM, especially if we need to serve many concurrent users with low latency. Unfortunately, cost, latency, and quality tradeoffs in such a high-concurrency, low-memory regime are rarely considered in the existing neural IR literature.

We tackle this with the following contributions. First, we present a methodology and benchmark for evaluating the concurrent serving of neural IR models under different traffic workloads and memory budgets. Second, we introduce \textbf{ColBERT-serve}\footnote{\url{https://github.com/stanford-futuredata/colbert-serve}}, which (1) incorporates a new memory-mapping architecture, permitting the bulk of ColBERTv2's index to reside \textit{on disk}, (2) \textbf{minimizes access to this index} via a multi-stage retrieval process, (3) \textbf{handles concurrent requests} in parallel with low latency and scales gracefully under load by adapting PISA and ColBERTv2, and (4) \textbf{preserves the quality of full ColBERTv2 retrieval} through a hybrid scoring technique. Third, we conduct an empirical evaluation that demonstrates the first ColBERT serving system that \textbf{can serve up to 4 queries per second on a server with as little as a few GBs of RAM} (90\% reduction in RAM usage for loading the model compared to the full ColBERTv2) for massive collections while preserving quality.

\section{Related Work}
Memory-mapping is a technique for accessing data from disk while only materializing accessed portions in memory \textit{on demand}. It is used in general-purpose approximate nearest neighbor search~\cite{johnson2019billion,Bernhardsson2023annoy}, some database systems~\cite{kotek2023mapdb}, and concurrently with our work also in neural IR~\cite{shrestha2023espn}. 
Memory-mapped indexes pose the challenge of minimizing the latency overhead incurred by page misses. Whereas \cite{shrestha2023espn} built a prefetcher to reduce the impact of SSD latencies, we seek to reduce the number of accesses to disk directly via multi-stage retrieval.
\cite{khattab2020colbert} studied the quality-cost tradeoff by using ColBERT to re-rank top 1000 results produced by BM25~\cite{robertson1995bm25}. This improved latency but comes at the cost of a substantial reduction in MRR and recall. 
\cite{mac2024repro} used ColBERTv2 to re-rank the candidates generated by BM25, and built a more effective system upon it using LADR~\cite{Kulkarni2023ladr}, though both systems led to quality loss on MS MARCO. 
We build a concurrent serving system for ColBERTv2 that permits the index to mostly reside on disk without sacrificing retrieval quality or latency under high traffic. We achieve this by leveraging a combination of memory-mapping and a multi-stage retrieval approach that utilizes scores from both the candidate generation and the re-ranking steps. This hybrid scoring method leverages the strengths of both stages, resulting in performance that can surpass fully in-memory ColBERTv2 retrieval.

\section{ColBERT-serve}

\paragraph{Memory-Mapped Storage} To deploy ColBERTv2 on memory-constrained machines, we introduce memory-mapping into the ColBERT implementation, specifically for the tensors encoding the compressed ColBERTv2 embeddings. This bypasses loading the index upfront and instead enables the operating system to manage limited memory resources, by bringing only accessed data into memory at the page granularity and evicting pages when RAM is insufficient. This reduces the RAM requirements by over $90\%$.

\paragraph{Concurrent Requests} We build a server-client architecture for deployment as well as experimentation for ColBERTv2. To this end, we improve ColBERTv2's multithreading compatibility by releasing Python's Global Interpreter Lock while invoking all underlying functionality of ColBERTv2 implemented as C++ extensions. Without this, multithreading for ColBERTv2 was prohibitively expensive as each query would block when extensions are invoked, so concurrency was only possible by launching multiple processes, which---without memory mapping---would scale memory consumption linearly with the number of processes. With support for memory mapping, we tune the number of threads used to serve each ColBERTv2 request and find that though multithreading improves performance under low load, single-threaded performance dominates under higher loads, hence we use only a single thread for all our experiments. In addition, we adapt the PISA~\cite{Antonio2019pisa} engine for this setting to support our server-client architecture with the multi-stage retrieval discussed next.

\paragraph{Multi-Stage Retrieval} Memory-mapping introduces a key challenge: due to the latency incurred by page misses, searching over MS MARCO with a memory-mapped index is approximately 2$\times$ slower than an in-memory index. 
We tackle this via a multi-stage ranking architecture, in which SPLADEv2~\cite{formal2021spladev2}, a sparse lexical and expansion model, serves as the first-stage retriever to minimize the number of documents we need to access from the ColBERT index. As a baseline,\footnote{We build on code from \url{https://github.com/stanford-futuredata/ColBERT}, \url{https://github.com/naver/splade}, and \url{https://github.com/pisa-engine/pisa}.} we use the standard ColBERTv2 with PLAID \cite{santhanam2022plaid} with a machine capable of fitting the entire index in memory. Then, we implement and study four different systems: (1) \textbf{MMAP ColBERTv2}, in which we apply memory-mapping to the end-to-end process of PLAID; (2) \textbf{SPLADEv2} w/ PISA, in which SPLADEv2 expands queries and the PISA engine performs efficient retrieval~\cite{Antonio2019pisa}; (3) \textbf{MMAP Rerank}, in which SPLADEv2 generates top-200 candidates per query and MMAP ColBERTv2 re-ranks them; and \textbf{MMAP Hybrid}, in which SPLADEv2's top-200 results are re-ranked via a linear interpolation between SPLADEv2 and MMAP ColBERTv2. For a given query $Q$ and document $D$, the hybrid score is given by:
\begin{equation*}
    S_{\text{hybrid}}(D,Q) = \alpha N(S_{\text{SPLADE}}(D,Q)) + (1-\alpha) N (S_{\text{ColBERT}}(D,Q))
\end{equation*}
where $S(*,*)$ is the score function, $N(*)$ is the normalization function, and $\alpha$ is a coefficient between 0 and 1. SPLADEv2 and ColBERTv2 produce scores of drastically different distributions, a likely source of quality for hybrid scoring. To combine these scores, we explored (1) linearly mapping each to the range of $[0,1]$, (2) min-max norm, and (3) z-norm. Among these, z-norm yielded the best results, so we select that as the normalization function, defined as: $N(x) = \frac{x-\Bar{x}}{S}$ where $\Bar{x}$ denotes the mean of samples and $S$ denotes the standard deviation.

\begin{table}[b]
  \centering
    \caption{AWS machine specifications}
    \begin{tabular}{lcccc}
    \hline
    &\textbf{\hspace{2mm}Control\hspace{2mm}} & \textbf{\hspace{2mm}MMAP MARCO\hspace{2mm}} &\textbf{\hspace{2mm}MMAP Wiki\hspace{2mm}} & \textbf{LoTTE}\\
    \hline
    AWS machine & r6a.4xlarge & m5ad.xlarge & r6id.xlarge  & c5ad.xlarge \\
    Disk Size (GB) & 950 & 150  & 237  & 150\\
    CPU Count & 16 & 4 & 4 & 4\\
    Memory (GB) & 128  & \textbf{16 (-88\%)} & \textbf{32 (-75\%)} & 8 \\
    Cost (\$/month) & 438 & \textbf{95 (-78\%)}  & \textbf{139 (-68\%)}  & 54 \\\hline
\end{tabular}
\label{table:specs}
\end{table}

\begin{table*}
  \centering
    \caption{
    Results of different methods on MS MARCO, Wikipedia (NQ-dev) and LoTTE (Lifestyle-dev) datasets. For SPLADEv2, we use the checkpoint of BT-SPLADE-L, from~\cite{Lassance2022effiency_splade}. We use a PISA index with a \texttt{block\_simdbp} encoding of block size 40 with a quantized scorer. This provides a highly optimized implementation for SPLADEv2. For MS MARCO, we report development results on Dev, on which we tune $\alpha$ for all datasets, and report evaluation results on the held-out evaluation set used by the ColBERT authors~\cite{khattab2020colbert,santhanam2022colbertv2}. For OOD datasets, we also find the optimal $\alpha$ (see in Table~\ref{tab:alpha}) and report the scores.}
    \begin{tabular}{cccc}
    \hline
    \textbf{Method}&\multicolumn{3}{c}{\textbf{MS MARCO Dev}}\\
    \hline
    &MRR@10&R@5&R@50\\
    \hline
    ColBERTv2 & 39.51 & 56.62 & 86.30 \\
    SPLADEv2 & 38.00 & 54.70& 85.04\\
     Rerank &39.50 & 56.65 & 86.64\\
     Hybrid ($\alpha=0.3)$& \textbf{40.22} & \textbf{57.38} & \textbf{86.98}\\
    \hline\\
  \end{tabular}
   \begin{tabular}{ccc}
    \hline
    \multicolumn{3}{c}{\textbf{MS MARCO 5K Test}}\\
    \hline
    MRR@10&R@5&R@50\\
    \hline
     40.57 & 57.78 & 86.14 \\
     38.62 & 54.92 & 84.84 \\
      40.55 & 57.78 & 86.36 \\
      \textbf{41.11} & \textbf{58.23} & \textbf{86.91}\\
    \hline\\
  \end{tabular}
  \begin{tabular}{ccccc}
    \hline
    \multicolumn{1}{c}{\textbf{Method}} & \multicolumn{2}{c}{\textbf{Wikipedia}} & \multicolumn{2}{c}{\textbf{LoTTE}} \\
    \hline
    &S@5&$\Delta$&S@5&$\Delta$\\
    \hline
    ColBERTv2 & \textbf{67.51} & & 74.6\\
    SPLADEv2 & 59.60 & -11.7\% & 70.7 & -5.6\%\\
    Rerank & 66.29  & -1.8\% & 74.3 & -0.4\%\\
    Hybrid ($\alpha=0.3$) & 65.78  & -2.6\% & 74.8 & +0.3\%\\
    Hybrid (optimal $\alpha$) & 66.34  & -1.7\% & \textbf{75.3} & \textbf{+0.9\%}\\
  \hline
  \end{tabular}
  \label{table:quality}
\end{table*}

\section{Evaluation}
We now test the impact of multi-stage retrieval on quality, of memory-mapping on RAM usage, and of both together on latency under varying traffic.

\paragraph{Methodology} We use MS MARCO Passage Ranking development set that contains 7K queries and 8.8M passages \cite{bajaj2016ms} as an ``in-domain'' benchmark for ColBERTv2 and SPLADEv2 and report MRR@10, Recall@5 and Recall@50. To test out-of-domain (OOD) generalization, we use Wikipedia Open-QA NQ-dev with 8.7K queries and 21M passages \cite{Kwiatkowski2019nq,lee2019latent} and LoTTE Search Lifestyle-dev with 417 queries and 269K passages \cite{santhanam2022colbertv2}. These popular datasets differ dramatically in size, with Wikipedia stressing RAM usage and LoTTE Lifestyle always fitting in memory. Following \cite{santhanam2022colbertv2}, we report Success@5. We report the latency at the 95th percentile observed by the concurrent clients in our client-server architecture under varying degrees of server load. We measure latency over the first 1K queries from each dataset, a sufficient size to saturate the system under high load conditions. The number of queries per second (QPS) is sampled using a Poisson distribution.
For MS MARCO and Wikipedia, we use an AWS \texttt{r6a.4xlarge} instance for the control (namely, the full ColBERTv2 baseline) experiment, and \texttt{m5ad.xlarge} and \texttt{r6id.xlarge} instances for SPLADEv2 and MMAP experiments, respectively. LoTTE Lifestyle's index is small enough to fit in a memory-restricted machine, so we run all experiments on a \texttt{c5ad.xlarge} instance. Key machine specifications are provided in Table \ref{table:specs}. 

\label{index}
\begin{figure}[h]
  \centering
  \includegraphics[width=0.55\textwidth]{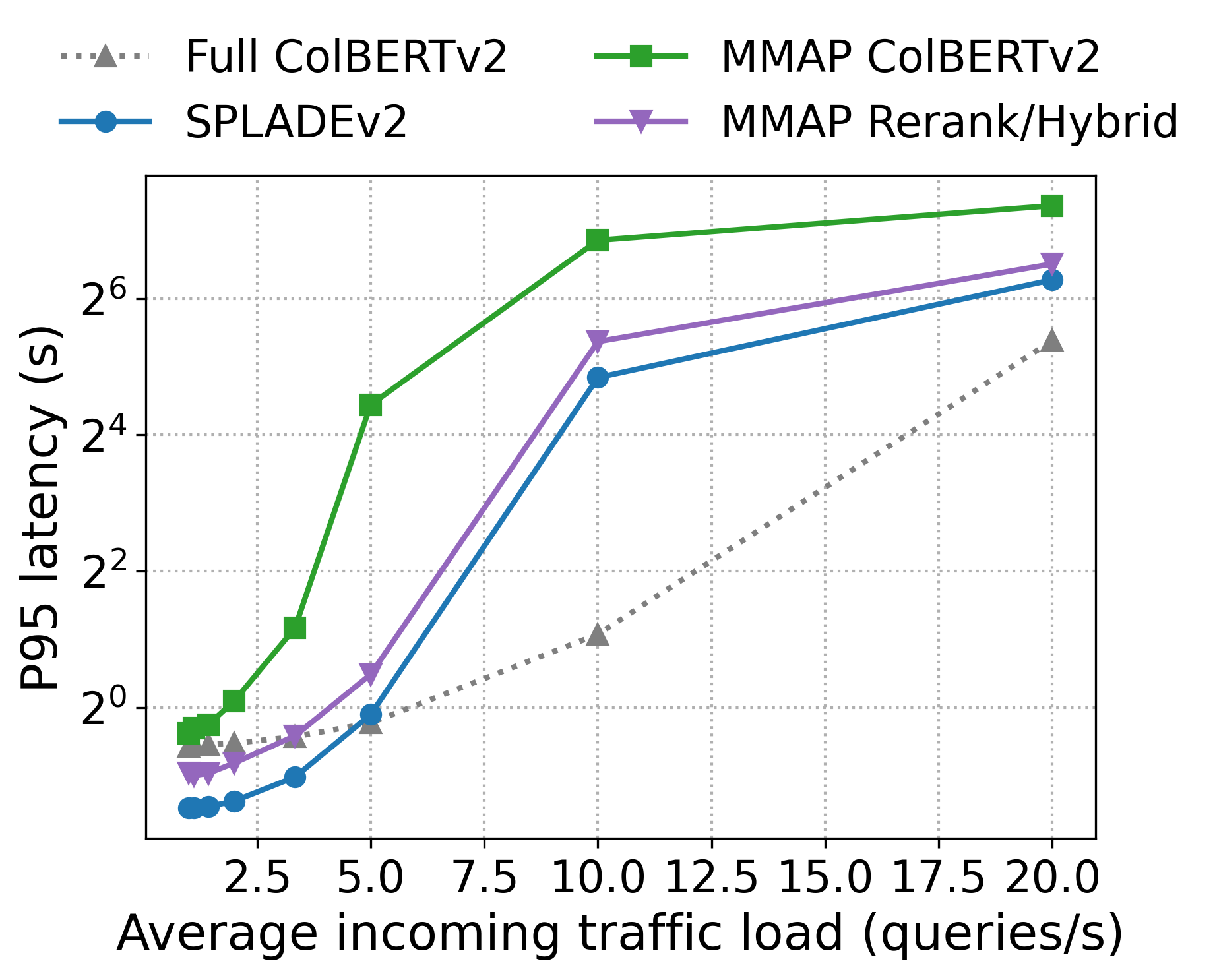}
  \caption{P95 Latency on Wikipedia Dataset.}
  \label{fig:wiki-latency-all}
\end{figure}

\begin{figure}[ht]
    \centering
    \begin{subfigure}{0.48\textwidth}
        \centering
        \includegraphics[width=\linewidth]{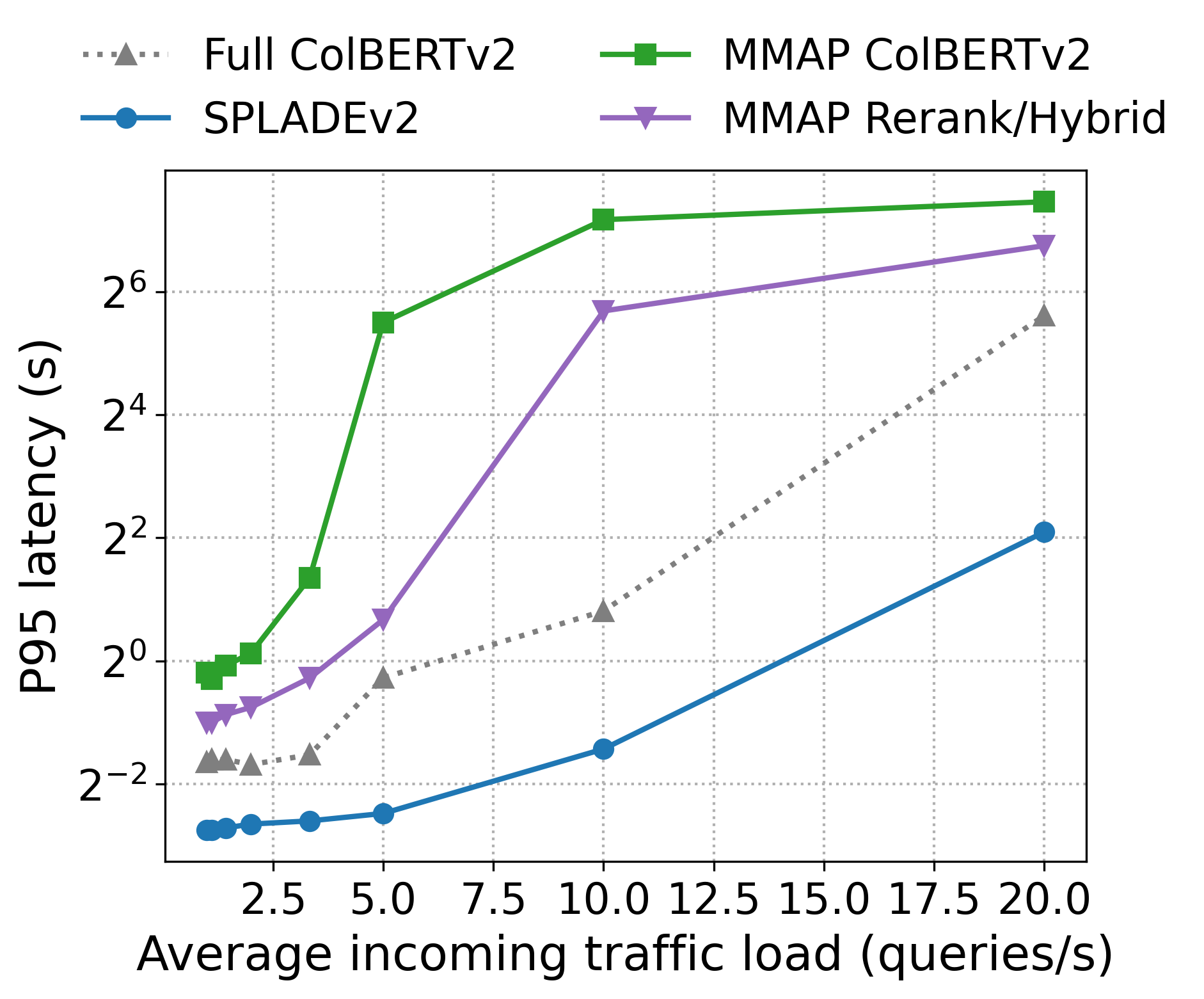}
        \caption{P95 Latency on MS MARCO}
        \label{fig:msmarco-latency-all}
    \end{subfigure}
    \hfill
    \begin{subfigure}{0.45\textwidth}
        \centering
        \includegraphics[width=\linewidth]{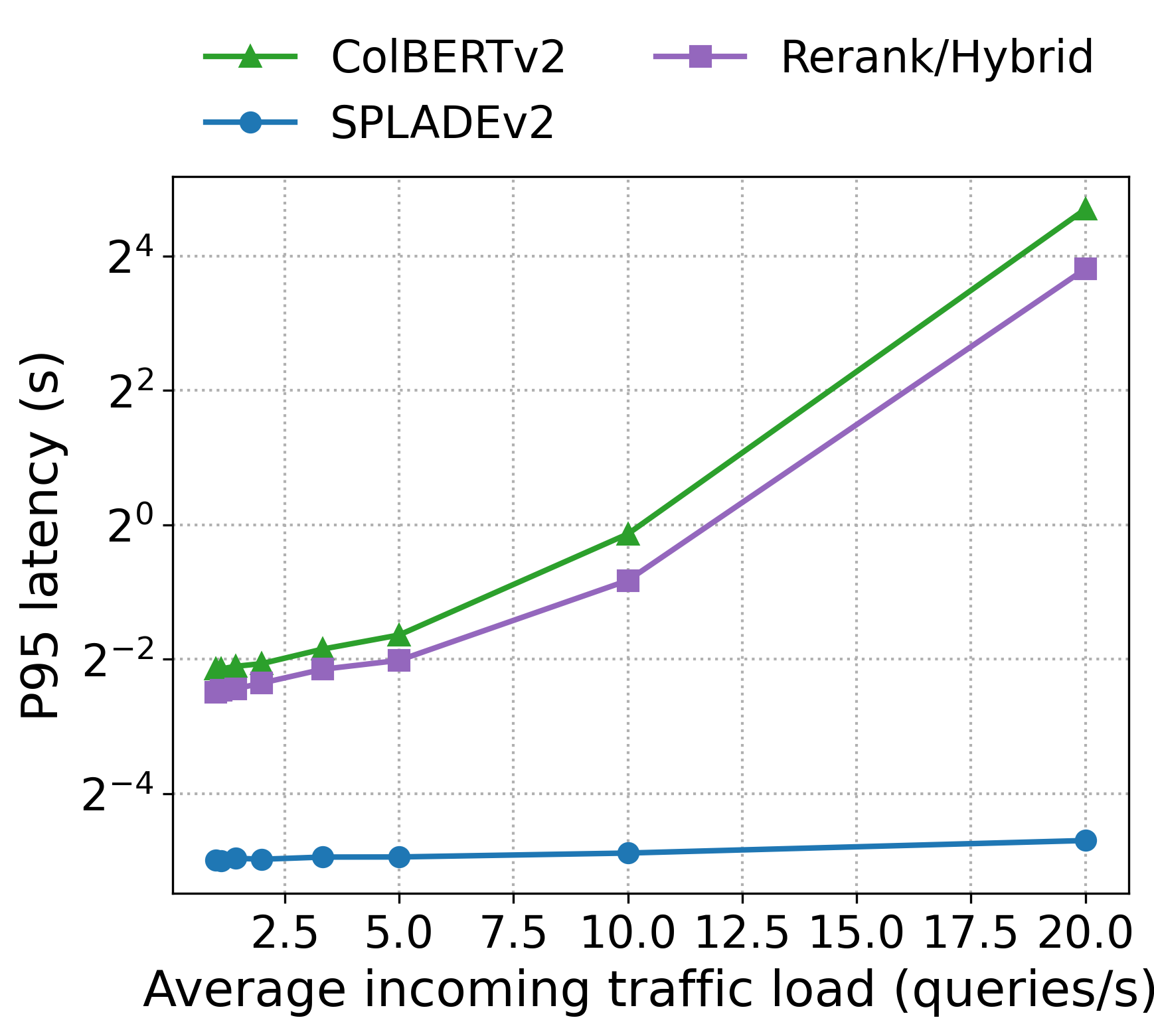}
        \caption{P95 Latency on LoTTE}
        \label{fig:lotte-latency-cpu}
    \end{subfigure}
    \caption{P95 Latency on MS MARCO and LoTTE. Note that full ColBERTv2 on MS MARCO is evaluated on a higher-end and more expensive machine (refer to Table~\ref{table:specs}) with a different physical processor, so its latency is only for reference and is not directly comparable to the MMAP methods.}
    \label{fig:mainfig}
\end{figure}

\paragraph{Retrieval Quality}
\label{quality}
Table \ref{table:quality} reports the quality of full ColBERTv2 scoring against more efficient approaches based on SPLADEv2, Rerank, and Hybrid scoring. We tune the parameter $\alpha$ for Hybrid on MS MARCO Dev and report the results of this setting (i.e., $\alpha = 0.3$) across all datasets. We can observe that Hybrid scoring is the most effective method on MS MARCO and that it outperforms the SPLADEv2 model and Rerank across every dataset. On Wikipedia, however, using a non-optimal $\alpha$ results in lower performance than Rerank. This suggests that tuning $\alpha$ on a held-out set can be important to OOD settings, though we leave this exploration for future work. Having confirmed the quality of the Rerank and especially Hybrid methods, we now proceed to evaluate the different efficiency dimensions.

\begin{table}
    \centering
    \caption{Retrieval quality of Hybrid changing with different $\alpha$. When $\alpha=0$, the method is equivalent to Rerank; when $\alpha=1$, the method is equivalent to SPLADEv2. Across all datasets, we see a similar trend where quality first increases then decreases.}
    \begin{tabular}{ccccccccccccc}
    \hline
         $\alpha$ & 0.0 & 0.1&0.2&0.3&0.4&0.5&0.6&0.7&0.8&0.9&1.0 \\\hline
         Wiki S@5 & 66.29 & \textbf{66.34} & 66.21 & 65.78 & 65.33 & 64.76 & 63.80 & 63.01 & 62.12 & 61.04 & 59.60\\\hline
         LoTTE S@5 & 74.3 & 74.1 & \textbf{75.3} & 74.8 & 74.6 & 74.3 & 73.4 & 72.4& 71.5 & 71.2 & 70.7\\\hline
         MARCO MRR@10 & 39.50 & 39.95 & 40.06 & \textbf{40.22} & 40.08 & 40.05 & 39.81 & 39.53 & 38.98 & 38.54 & 38.00&\\
         \hline
    \end{tabular}
    \label{tab:alpha}
\end{table}

\paragraph{RAM Usage}
We measure memory usage for loading ColBERTv2 on MS MARCO and Wikipedia by recording the difference in RSS memory before and after loading. For the memory-mapped approaches, only the model checkpoint and index metadata are loaded into memory, resulting in a substantial reduction of RAM usage, by $90\%$ for MS MARCO (from 23.4 GB to 2.3 GB) and $92\%$ for Wikipedia (from 98.3 GB to 8.2 GB). Our approach allows us to host the indexes on machines with significantly lower RAM capacities, and reduces machine cost by 78\% for MS MARCO and 68\% for Wikipedia, as shown in Table \ref{table:specs}.

\paragraph{Latency on Varying Traffic}
\label{res:latency}

Figure \ref{fig:wiki-latency-all} compares latency across methods on Wikipedia. The exceedingly optimized PISA implementation of SPLADEv2, using the efficiency-optimized BT-SPLADE-L model checkpoint~\cite{Lassance2022effiency_splade}, has the lowest latency, though this comes at the steep reduction in quality, especially out of domain, presented earlier.
Next, although the Rerank/Hybrid methods incur higher latency than SPLADE, they are markedly faster than the memory-mapped ColBERTv2 method. The Rerank/Hybrid methods maintain low latency with QPS up to $1/0.2=5$ queries per second. When QPS exceeds this, the system is saturated, leading to a sharper increase in latency due to queuing time. Note that as shown in Table \ref{table:specs}, full ColBERTv2 experiments were conducted on a more expensive machine that fits the index in RAM, for reference. Despite this, the Rerank/Hybrid methods still achieve lower latency than full ColBERTv2 on QPS $<1/0.3=3.3$, highlighting the value of multi-stage retrieval.
Figure \ref{fig:msmarco-latency-all} shows similar trends on MS MARCO, where our Rerank/Hybrid systems greatly reduce the latency of memory-mapping ColBERTv2 across every traffic load. Note that full ColBERTv2 is evaluated on a machine with a different physical processor, so its latency is only for reference and is not directly comparable to the memory-mapped methods. Lastly, Figure \ref{fig:lotte-latency-cpu} reports very similar patterns for for LoTTE. Note that we do not apply memory mapping for LoTTE, whose ColBERTv2 index fits easily in the RAM of our smallest machines.

\section{Conclusion}
We presented a serving system based on memory-mapping, hybrid scoring, and support for concurrent requests. We introduced an evaluation methodology for assessing the neural IR tradeoffs in the concurrent, memory-constrained regime and demonstrated the first ColBERT serving system that can serve several queries per second on a server with as little as a few GBs of RAM.

\section*{Acknowledgements}
This preprint has not undergone peer review (when applicable) or any post-submission improvements or corrections. The Version of Record of this contribution is published in Advances in Information Retrieval. ECIR 2025. Lecture Notes in Computer Science, vol 15575, and is available online at \url{https://doi.org/10.1007/978-3-031-88717-8_3}.

% \bibliographystyle{splncs04}
% \bibliography{ref}

\begin{thebibliography}{10}
\providecommand{\url}[1]{\texttt{#1}}
\providecommand{\urlprefix}{URL }
\providecommand{\doi}[1]{https://doi.org/#1}

\bibitem{bajaj2016ms}
Bajaj, P., Campos, D., Craswell, N., Deng, L., Gao, J., Liu, X., Majumder, R., McNamara, A., Mitra, B., Nguyen, T., et~al.: Ms marco: A human generated machine reading comprehension dataset. arXiv preprint arXiv:1611.09268  (2016)

\bibitem{Bernhardsson2023annoy}
Bernhardsson, E.: Spotify/annoy: Approximate nearest neighbors in c++/python optimized for memory usage and loading/saving to disk, \url{https://github.com/spotify/annoy}

\bibitem{faysse2024colpaliefficientdocumentretrieval}
Faysse, M., Sibille, H., Wu, T., Omrani, B., Viaud, G., Hudelot, C., Colombo, P.: Colpali: Efficient document retrieval with vision language models (2024), \url{https://arxiv.org/abs/2407.01449}

\bibitem{formal2021spladev2}
Formal, T., Lassance, C., Piwowarski, B., Clinchant, S.: {SPLADE} v2: Sparse lexical and expansion model for information retrieval. CoRR  \textbf{abs/2109.10086} (2021), \url{https://arxiv.org/abs/2109.10086}

\bibitem{hofstatter2022colberter}
Hofst\"{a}tter, S., Khattab, O., Althammer, S., Sertkan, M., Hanbury, A.: Introducing neural bag of whole-words with colberter: Contextualized late interactions using enhanced reduction. In: Proceedings of the 31st ACM International Conference on Information \& Knowledge Management. p. 737–747. CIKM '22, Association for Computing Machinery, New York, NY, USA (2022). \doi{10.1145/3511808.3557367}, \url{https://doi.org/10.1145/3511808.3557367}

\bibitem{johnson2019billion}
Johnson, J., Douze, M., J{\'e}gou, H.: Billion-scale similarity search with gpus. IEEE Transactions on Big Data  \textbf{7}(3),  535--547 (2019)

\bibitem{khattab2020colbert}
Khattab, O., Zaharia, M.: Colbert: Efficient and effective passage search via contextualized late interaction over bert. In: Proceedings of the 43rd International ACM SIGIR Conference on Research and Development in Information Retrieval. p. 39–48. SIGIR '20, Association for Computing Machinery, New York, NY, USA (2020). \doi{10.1145/3397271.3401075}, \url{https://doi.org/10.1145/3397271.3401075}

\bibitem{kotek2023mapdb}
Kotek, J.: Jankotek/mapdb: Mapdb provides concurrent maps, sets and queues backed by disk storage or off-heap-memory. it is a fast and easy to use embedded java database engine., \url{https://github.com/jankotek/mapdb/}

\bibitem{Kulkarni2023ladr}
Kulkarni, H., MacAvaney, S., Goharian, N., Frieder, O.: Lexically-accelerated dense retrieval. In: Proceedings of the 46th International ACM SIGIR Conference on Research and Development in Information Retrieval. p. 152–162. SIGIR '23, Association for Computing Machinery, New York, NY, USA (2023). \doi{10.1145/3539618.3591715}, \url{https://doi.org/10.1145/3539618.3591715}

\bibitem{Kwiatkowski2019nq}
Kwiatkowski, T., Palomaki, J., Redfield, O., Collins, M., Parikh, A., Alberti, C., Epstein, D., Polosukhin, I., Devlin, J., Lee, K., Toutanova, K., Jones, L., Kelcey, M., Chang, M.W., Dai, A.M., Uszkoreit, J., Le, Q., Petrov, S.: {Natural Questions: A Benchmark for Question Answering Research}. Transactions of the Association for Computational Linguistics  \textbf{7},  453--466 (08 2019). \doi{10.1162/tacl_a_00276}, \url{https://doi.org/10.1162/tacl\_a\_00276}

\bibitem{Lassance2022effiency_splade}
Lassance, C., Clinchant, S.: An efficiency study for splade models. In: Proceedings of the 45th International ACM SIGIR Conference on Research and Development in Information Retrieval. p. 2220–2226. SIGIR '22, Association for Computing Machinery, New York, NY, USA (2022). \doi{10.1145/3477495.3531833}, \url{https://doi.org/10.1145/3477495.3531833}

\bibitem{lee2019latent}
Lee, K., Chang, M.W., Toutanova, K.: Latent retrieval for weakly supervised open domain question answering. In: Proceedings of the 57th Annual Meeting of the Association for Computational Linguistics. Association for Computational Linguistics (2019)

\bibitem{mac2024repro}
MacAvaney, S., Tonellotto, N.: A reproducibility study of plaid. In: Proceedings of the 47th International ACM SIGIR Conference on Research and Development in Information Retrieval. p. 1411–1419. SIGIR '24, Association for Computing Machinery, New York, NY, USA (2024). \doi{10.1145/3626772.3657856}, \url{https://doi.org/10.1145/3626772.3657856}

\bibitem{Antonio2019pisa}
Mallia, A., Siedlaczek, M., Mackenzie, J., Suel, T.: {PISA:} performant indexes and search for academia. In: Proceedings of the Open-Source {IR} Replicability Challenge co-located with 42nd International {ACM} {SIGIR} Conference on Research and Development in Information Retrieval, OSIRRC@SIGIR 2019, Paris, France, July 25, 2019. pp. 50--56 (2019), \url{http://ceur-ws.org/Vol-2409/docker08.pdf}

\bibitem{robertson1995bm25}
Robertson, S., Walker, S., Jones, S., Hancock-Beaulieu, M.M., Gatford, M.: Okapi at trec-3. In: Overview of the Third Text REtrieval Conference (TREC-3). pp. 109--126. Gaithersburg, MD: NIST (January 1995), \url{https://www.microsoft.com/en-us/research/publication/okapi-at-trec-3/}

\bibitem{santhanam2022plaid}
Santhanam, K., Khattab, O., Potts, C., Zaharia, M.: Plaid: an efficient engine for late interaction retrieval. In: Proceedings of the 31st ACM International Conference on Information \& Knowledge Management. pp. 1747--1756 (2022)

\bibitem{santhanam2022colbertv2}
Santhanam, K., Khattab, O., Saad-Falcon, J., Potts, C., Zaharia, M.: {C}ol{BERT}v2: Effective and efficient retrieval via lightweight late interaction. In: Carpuat, M., de~Marneffe, M.C., Meza~Ruiz, I.V. (eds.) Proceedings of the 2022 Conference of the North American Chapter of the Association for Computational Linguistics: Human Language Technologies. pp. 3715--3734. Association for Computational Linguistics, Seattle, United States (Jul 2022). \doi{10.18653/v1/2022.naacl-main.272}, \url{https://aclanthology.org/2022.naacl-main.272}

\bibitem{shrestha2023espn}
Shrestha, S., Reddy, N., Li, Z.: Espn: Memory-efficient multi-vector information retrieval. In: Proceedings of the 2024 ACM SIGPLAN International Symposium on Memory Management. p. 95–107. ISMM 2024, Association for Computing Machinery, New York, NY, USA (2024). \doi{10.1145/3652024.3665515}, \url{https://doi.org/10.1145/3652024.3665515}

\bibitem{thakur2021beir}
Thakur, N., Reimers, N., R{\"u}ckl{\'e}, A., Srivastava, A., Gurevych, I.: Beir: A heterogenous benchmark for zero-shot evaluation of information retrieval models. arXiv preprint arXiv:2104.08663  (2021)

\end{thebibliography}

\end{document}